\documentclass[12pt]{article}

\usepackage{graphicx}
\usepackage{cite}
\usepackage{amsmath}

\begin{document}
\title{EFFECT OF IONIC ORDERING IN CONDUCTIVITY EXPERIMENTS OF DNA AQUEOUS SOLUTIONS}
\author{O.O. LIUBYSH$^{1}$, O.M. ALEKSEEV$^{1}$, S.Yu. TKACHOV$^{1}$, \\S.M.
PEREPELYTSYA$^{2}$\\
$^{1}$Taras Shevchenko National University of Kyiv,\\ 64,
Volodymyrska Str., Kyiv 01033, Ukraine
\\ $^{2}$Bogolyubov Institute
for Theoretical Physics, NAS of Ukraine,\\14-b Metrolohichna Str.,
Kiev, 03680, Ukraine \\ perepelytsya@bitp.kiev.ua} \maketitle

\setcounter{page}{1}%
\maketitle

\begin{abstract}
The effects of ionic ordering in DNA water solutions are studied
by conductivity experiments. The conductivity measurements are
performed for the solutions of DNA with KCl salt in the
temperature range from 28 to 70~$^{0}$C. Salt concentration vary
from 0 to 2~M. The conductivity of solutions without DNA but with
the same concentration of KCl salt are also performed. The results
show that in case of salt free solution of DNA the melting process
of the double helix is observed, while in case of DNA solution
with added salt the macromolecule denaturation is not featured.
For salt concentrations lower than some critical one (0.4~M) the
conductivity of DNA solution is higher than the conductivity of
KCl water solution without DNA. Starting from the critical
concentration the conductivity of KCl solution is higher than the
conductivity of DNA solution with added salt. For description of
the experimental data phenomenological model is elaborated basing
on electrolyte theory. In framework of the developed model a
mechanism of counterion ordering is introduced. According to this
mechanism under the low salt concentrations electrical
conductivity of the system is caused by counterions of DNA
ion-hydrate shell. Increasing the amount of salt to the critical
concentration counterions condense on DNA polyanion. Further
increase of salt concentration leads to the formation of DNA-salt
complexes that decreases the conductivity of the system.
\end{abstract}

\section{Introduction}
\label{intro} DNA double helix is a strong polyelectrolyte which
in aqueous solutions dissociates into macromolecular polyanion and
mobile cations (counterions) \cite{1,2}. Under the natural
conditions the counterions are positively charged metal ions
(usually Na$^{+}$ or K$^{+}$) that neutralize negatively charged
phosphate groups of macromolecule backbone. The counterions and
water molecules form an ion-hydrate shell around DNA stabilizing
the structure of the double helix \cite{3,4,5,6,7,8}. In spite of
significant mobility of counterions they are organized as the
dynamical structure around macromolecule. This structure may be
rather regular due to the homogeneity of DNA backbone \cite{9,10}.
The ordering of counterions around DNA macromolecule determines
the elastic properties of the double helix (bending, twisting,
denaturation), DNA interaction with biologically active compounds
(proteins, drugs), and compaction mechanisms of macromolecule in
small volumes (chromosomes, viral capsids)
\cite{11,12,13,14,15,16}. The study of dynamical ordering of DNA
counterions is of paramount importance for understanding the
mechanisms of DNA biological functioning.

Effects of dynamical ordering of counterions around DNA double
helix may become apparent in conductivity experiments due to the
interaction of charged particles of the solution with the electric
field. As known the electric current in DNA water solutions is
caused by the motion of counterions and DNA macromolecules
\cite{17,18,19,20,21,22,23}. In case of DNA solutions without
added salt (salt free solution) the conductivity increases as the
concentration of DNA increases because of counterion dynamics in
the ion-hydrate shell of macromolecule \cite{17,18}. Heating the
system the conductivity gradually increases, and under the
temperatures of the double helix melting
 there is a sudden change of conductivity that is caused
by intensive ejection of counterions from DNA ion-hydrate shell
\cite{18}. In case of DNA solution with added salt the
conductivity of the system depends on the both counterion type and
salt concentration \cite{17,18}. The dependence on counterion type
is caused by different electrophoretic mobility of ions mostly
\cite{17}, while the dependence on salt concentration may reflect
the ordering of the ions in solution. The experimental data show
that under the low concentration of added NaCl salt the
conductivity of DNA solution is higher than the conductivity of
NaCl electrolyte solution, but starting from some defined
concentration the conductivity of DNA solution becomes lower than
the conductivity of electrolyte \cite{19}. The reason of such
concentration dependence of conductivity of DNA solutions is not
determined yet.

To elucidate the microscopic picture of conductivity process in
DNA solution the phenomenological approaches have been developed,
and atom-atom calculations have been performed \cite{24,25,26,27}.
The results have been showed that the dynamics of counterions in
close vicinity to DNA surface is modulated by the charged atomic
groups of the double helix backbone. Part of the time counterions
spend in complex with DNA (about 1~ns) and another part in free
state \cite{28,29,30,31}. Free counterions determine the
conductivity of DNA solution in many respects that has been taken
into consideration in phenomenological models \cite{24,26}. In the
same time, the counterions, tethered to the phosphate groups, form
ordered dynamical structure along DNA backbone that may be
considered as the lattice of ionic type (ion-phosphate lattice)
\cite{9,10}. The existence of the ion-phosphate lattice is
confirmed by observing the modes of ion-phosphate vibrations in
the low-frequency Raman spectra of DNA ($<200$ cm$^{-1}$)
\cite{32,33,34,35}. The ordering of counterions around the double
helix and the formation of the ion-phosphate lattice should affect
the conductivity of DNA water solutions.

The goal of the present work is to study the manifestations of
counterion ordering around DNA double helix in conductivity
experiments of DNA water solutions with added salt. To solve this
problem the conductivity of DNA water solutions with KCl salt is
studied experimentally in the Section 2. As the result the
concentration dependence (0$\div$2M) of conductivity of DNA
solutions is obtained at temperature range from 28 to 70 $^{0}$C
(Section~\ref{res}). For the interpretation of experimental data
the phenomenological model is developed basing on  electrolyte
theory (Section~\ref{model}). In the Section~\ref{disc} possible
mechanism of ionic ordering around DNA double helix is discussed.

\section{Materials and Methods}
\label{materials} The samples have been prepared using sodium salt
of DNA from salmon testes purchased from Sigma-Aldrich Company
(product number D1626). The average length of DNA macromolecules
is about 2000 base pairs \cite{36}. To prepare the samples of DNA
water solutions the powder of DNA has been diluted in deionized
water to the concentration 10~mg/ml.  To decrease the viscosity of
DNA solution it has been treated by the laboratory automatic mixer
and than cooled to the $0~^{0}$C without freezing of water. Than
the initial solution has been diluted to 2~mg/ml concentration of
DNA, and KCl salt has been added to this solution.  The
concentrations of added salt in the obtained solutions are as
follows: 0.4, 0.8, 1.2, 1.6, and 2~M. Water solutions without DNA
but with the same concentrations of KCl salt have been also
prepared. As the result two series of the samples have been
prepared: KCl electrolyte solutions, and water solutions of DNA
with KCl salt.

\begin{figure}[t!]
\begin{center}
\resizebox{0.35\textwidth}{!}{%
  \includegraphics{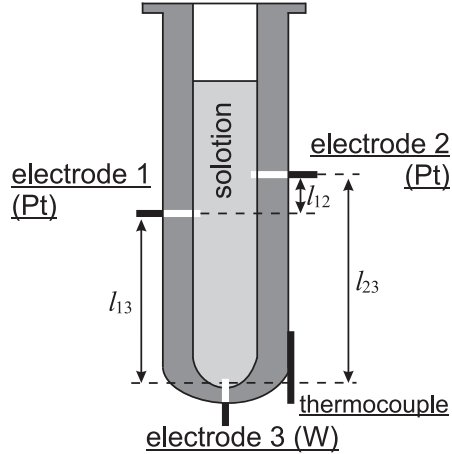}
}
\end{center}
%\vspace{5cm}       % Give the correct figure height in cm
\caption{Scheme of the experimental capillary cell.}
\label{scheme}      % Give a unique label
\end{figure}

To measure the resistance of the sample the solution (about
0.3~ml) is poured into cylindrical capillary made of quartz glass
with two platinum electrodes (electrode 1 and electrode 2) and one
wolfram electrode (electrode 3) incorporated into the capillary
walls (Fig. \ref{scheme}). The experimental cell is placed into
thermostat. The resistance has been determined by alternating
current with the frequency 80~MHz.

The measured resistance has the contributions from polarization of
the sample and electrodes. To exclude the electrode contribution
the measurements have been performed for different pairs of
electrodes: 1 and 3, 2 and 3 (Fig. \ref{scheme}). The resistance
in this case may be presented as follows:
\begin{equation}
R_\text{13} = \frac{l_\text{13}}{\pi r^{2} \sigma} + R_\text{Pt} +
R_{W}; \label{R13}
\end{equation}
\begin{equation}
R_\text{23} = \frac{l_\text{23}}{\pi r^{2} \sigma} + R_\text{Pt} +
R_{W}; \label{R23}
\end{equation}
where \(R_\text{13}\) and \(R_\text{23}\) are measured resistances
between electrodes 1 and 3, 2 and 3, respectively; \(l_\text{13}\)
and \(l_\text{23}\) are the distances between electrodes 1-3 and
2-3, respectively; \textit{r} is the capillary radius; \(\sigma \)
is the specific conductance, \(R_\text{Pt}\) and \(R_\text{W}\)
are the resistances of platinum and wolfram electrodes,
respectively. The first terms in equations (\ref{R13}) and
(\ref{R23}) describe the resistance of the sample, while the
second and the third terms describe the polarization resistance of
electrodes. The difference of formulae (\ref{R13}) and (\ref{R23})
gives the following formula for conductivity of the sample:
\begin{equation}
\sigma = \frac{l_\text{12}}{\pi r^{2} (R_\text{13}-R_\text{23})},
\label{sigma3}
\end{equation}
where $l_\text{12}$ is the distances between electrodes 1-2. Using
the formula (\ref{sigma3}) the conductivity of the samples are
determined.

\section{Results}
\label{res} 
\begin{figure}
\begin{center}
\resizebox{0.5\textwidth}{!}{%
  \includegraphics{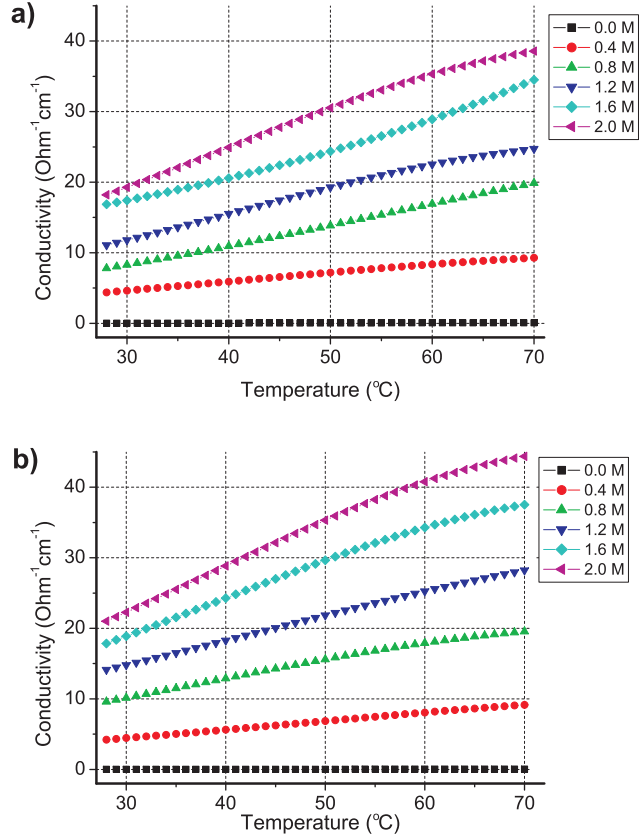}
}
\end{center}
 \caption{Temperature dependence of electrical conductivity of the samples. a) DNA solution with added KCl salt. b) KCl solution.}
\label{sigma(T)}      % Give a unique label
\end{figure}

The temperature dependences of electrical conductivity
of salt solution ($\sigma_\text{KCl}$) and solutions of DNA with
added salt ($\sigma_\text{DNA+salt}$) are obtained
(Fig.~\ref{sigma(T)}). The results show that the conductivity of
the samples increases as the temperature increases for all
considered samples.

According to the activation mechanism of ion motion in the
solution the temperature dependence of conductivity of the system
may be considered analogically to the Arrhenius equation for the
temperature dependence of chemical reaction rates \cite{37}:
\begin{equation}
\sigma = \sigma_\text{0} \exp\left(- \frac{\Delta
E}{k_\text{B}T}\right), \label{sigma4}
\end{equation}
where $\sigma_\text{0}$ is the coefficient; $\Delta E$ is the
potential barrier; \(k_\text{B}\) is the Boltzmann constant; \(T\)
is the temperature. The exponent describes the probability of
ionic jumping over the potential barrier due to the thermal
fluctuations. To analyze the temperature dependence of electrical
conductivity let us use the Arrenius coordinates, describing the
logariphm of conductivity as the function of transverse
temperature. From formula (\ref{sigma4}) it is seen that the
temperature dependence of conductivity in Arrhenius coordinates
should be linear.

\begin{figure}
\begin{center}
\resizebox{0.5\textwidth}{!}{%
  \includegraphics{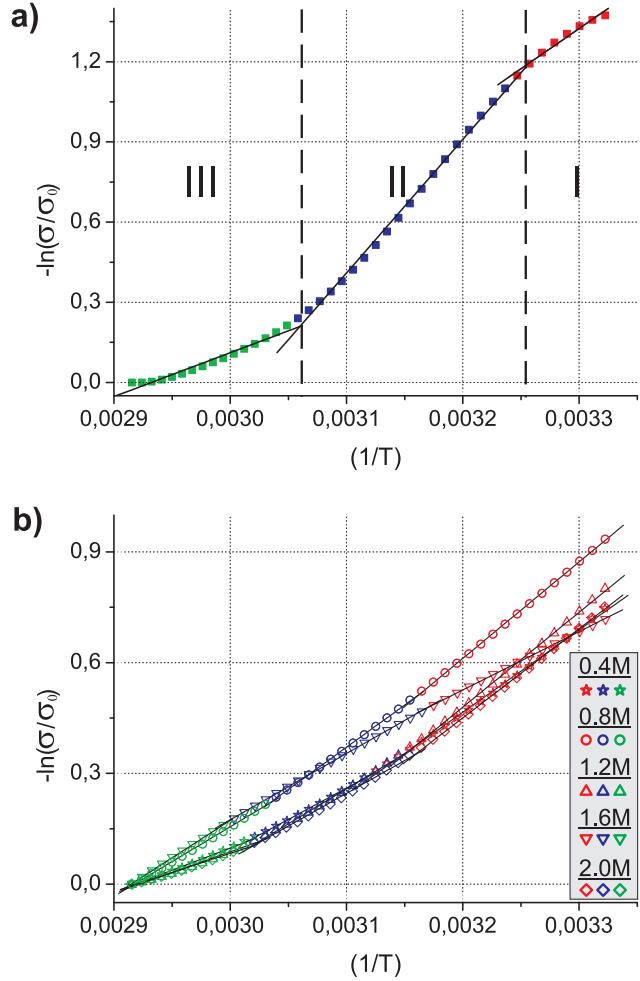}
}
\end{center}
 \caption{The Arrhenius plot for DNA water solutions. a)
Salt free solution. I is range of double-stranded DNA (red
points); II is transition range of double helix melting (blue
points); III is range of single-stranded DNA (green points). b)
Solution of DNA with added salt. Solid black lines is the linear
approximation.}

\label{Arr}      % Give a unique label
\end{figure}

The Arrhenius plot for salt free solution of DNA
(Fig.~{\ref{Arr}a}) shows that there are two braking points
separating distinguishable linear ranges. Linear ranges in the
Arrhenius plot characterize melting process of DNA double helix
\cite{18}. The range I (28$\div$37~$^{0}$C) and III
(54$\div$70~$^{0}$C) features the double stranded and
single-stranded DNA, respectively. The range II
(37$\div$54~$^{0}$C) characterizes denaturation of DNA
macromolecules. In case of DNA with added salt the difference
between linear ranges in the Arrhenius plot is not prominent and
braking points are hardly distinguishable (Fig.~{\ref{Arr}b}). The
influence of added salt may be explained by additional
neutralization of the negatively charged atomic groups of the
double helix by salt ions.

Different ranges in Arrhenius plot characterizes different
activation energy of ionic motion in the solution. The values of
potential barrier \(\Delta E\) are determined as a slope of the
lines in Fig.~{\ref{Arr}} (Table~{\ref{barrier}}). The results
show that in salt free solution of DNA before the melting
temperature (range I) the activation energy is rather large
comparing to the electrolyte solution (about 25~kJ/mole). In the
transition range (range II) the activation energy (about
43~kJ/mole) increases almost twice comparing to the range I that
is effectively caused by the ejection of counterions from DNA
ion-hydrate shell \cite{9}. Under the melting temperatures (range
III) \(\Delta E\) values decrease.

\begin{table}
\noindent\caption{Values of potential barrier $\Delta E$ for the
ion motion in DNA solution (kJ/mole).}\vskip3mm\tabcolsep4.5pt
%values of barrier
\begin{center}
\noindent{\footnotesize
\begin{tabular}{lcccccccc}
\hline
   & \multicolumn{1}{c}{0~M}
   & \multicolumn{1}{c}{0.4~M}
   & \multicolumn{1}{c}{0.8~M}
   & \multicolumn{1}{c}{1.2~M}
   & \multicolumn{1}{c}{1.6~M}
   & \multicolumn{1}{c}{2.0~M}
   & \multicolumn{1}{c}{Mean} \\
\hline
I & 25.01 & 18.55 & 21.71 & 21.33 & 13.15 & 20.10 & $19 \pm 4$ \\
II & 41.52 & 14.07 & 18.99 & 15.78 & 14.90 & 14.74 & $16 \pm 2$\\
III & 13.62 & 9.98 & 15.65 & 9.36 & 16.83 & 8.61 & $12 \pm 4$ \\
 &   &   &   &   &   &   & $16 \pm 4$ \\\hline
\end{tabular}
}
\end{center}
 \label{barrier}
\end{table}

In the solutions of DNA with added salt the potential barriers
\(\Delta E\) of different ranges are rather close. Comparing to
the salt free solution \(\Delta E\) values only slightly decrease
in case of the range I and range III, while in case of the range
II it decrease more than twice. The fact of comparatively low
activation barrier in the range II indicates that the added salt
increases the melting temperature of DNA double helix that is also
observed in calorimetric experiments \cite{2}.

\begin{figure}[b!]
\begin{center}
\resizebox{0.5\textwidth}{!}{%
  \includegraphics{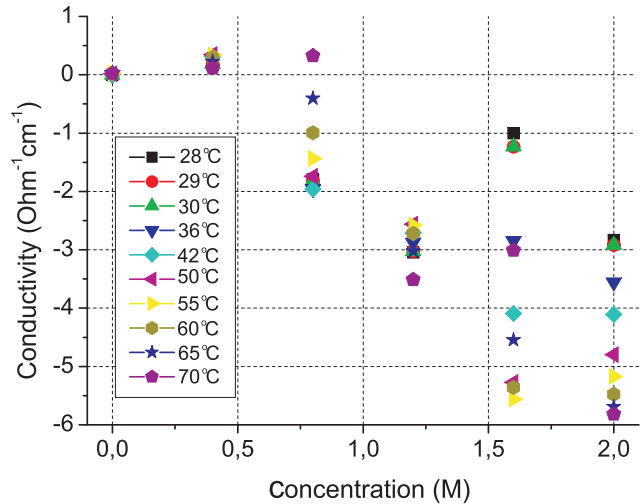}
}
\end{center}
 \caption{Concentration dependence for difference between conductivities of DNA and electrolyte solutions.}
\label{fig4}       % Give a unique label
\end{figure}

Increasing the concentration of added salt the conductivity of the
both DNA solution and electrolyte increases (Fig.~\ref{sigma(T)}).
To compare the conductivity of DNA solution with added salt and
the conductivity of KCl electrolyte solution the difference
\(\Delta\sigma~=~\sigma_\text{DNA+KCl}~-~\sigma_\text{KCl}\) is
analyzed (Fig.~\ref{fig4}). The results show that at
concentrations lower than some critical one (about \(c_\text{cr}
\approx \) 0.4~M) the conductivity of DNA solution is higher than
the conductivity of salt solution
(\(\sigma_\text{DNA+salt}~>~\sigma_\text{salt} \)). Under the
critical concentration (\(c = c_\text{cr}\)) the conductivity of
DNA solution and KCl solution are equal (\(\Delta\sigma \)~=~0).
Starting from critical concentration (\(c~>~c_\text{cr}\)) the
conductivity of DNA solution is lower than the conductivity of
respective electrolyte (\(\sigma_\text{DNA+salt} <
\sigma_\text{salt}\)). The dependence of conductivity of DNA
solution on salt concentration at range from 0 to \(c_\text{cr}\)
is about the same for different temperatures, while at the
concentration range from $c_\text{cr}$ to 2~M it is different for
different temperatures. The changes of $\Delta \sigma$ values
should reflect the structure changes in DNA solution.

\section{Model}
\label{model} To understand the mechanism of electrical
conductance of DNA water solution let us analyze the state of DNA
macromolecule in the solution. Due to the large contour length DNA
macromolecules are coiled shaped. The size of DNA coils may be
estimated with the use of the persistence model \cite{2,38}. In
framework of this model the root-mean-square distance between the
ends of macromolecule is determined as follows:
\begin{equation}
\bar{D^{2}} = 2P^{2}(L/P - 1 + e^{-L/P}), \label{model5}
\end{equation}
where \(L\) and \(P\) are the contour and persistence lengths of
macromolecule, respectively. The contour length for DNA from
salmon testes is \(L \approx \)~0.68~$\mu$m \cite{36}. The
persistence length of DNA is \(P \approx \)~500~A \cite{3,36}.
Using such parameters the average volume of DNA coils is estimated
0.02~$\mu$m$^{3}$. Taking into consideration that average number
of DNA macromolecules in 1~ml of the experimental solution is
10$^{15}$ the total volume of DNA coils should be about 20~ml. One
can conclude that macromolecule coils overlap in the considered
solution and the conductivity process may be determined by mobile
ions only, because the migration of single DNA macromolecules is
labored.

The number of mobile ions involved in conductivity process is
determined by the concentration of DNA counterions and ions of
added salt. Taking this into consideration the conductivity of DNA
solution may be presented as follows:
\begin{equation}
\sigma_\text{DNA+salt}(c) = \sigma_\text{1}(c) +
\sigma_\text{2}(c), \label{model6}
\end{equation}
where \(\sigma_\text{1}(c)\) is the conductivity determined by the
motion of salt ions (bulk ions); \(\sigma_\text{2}(c)\) is the
conductivity determined by mobility of counterions in ion-hydrate
shell of DNA; \(c\) is the equivalent concentration of added salt.

Taking into consideration that salt ions may condense on DNA
macromolecule the conductivity of bulk ions may be considered as
follows:
\begin{equation}
\sigma_\text{1}(c) = \sigma_\text{salt}(c) -
A_\text{1}(c)(\lambda^{+} + \lambda^{-}) \label{model7},
\end{equation}
where \(\sigma_\text{salt}(c)\) is the contribution of salt ions
to the conductivity of the system; \( A_\text{1}(c)\) is the
concentration of salt ions condensed on DNA macromolecule;
\(\lambda^{+}\) and \(\lambda^{-}\) are equivalent mobility of
positively and negatively charged ions, respectively. The second
term in (\ref{model7}) describes the conductivity decrease caused
by the association of the positively and negatively charged ions
with DNA. Note that the negatively charged ions may associate with
the positively charged ions that are already tethered to the
phosphate groups of DNA backbone.

The contribution from DNA counterions to the conductivity of the
system may be taken into consideration as follows:
\begin{equation}
\sigma_\text{2}(c) = c_\text{p} \lambda^{+} - A_\text{2}(c)
\lambda^{+}, \label{model8}
\end{equation}
where \(c_\text{p}\) is the concentration of DNA counterions that
approximately equals to the number of DNA phosphate groups;
\(A_\text{2}(c)\) is the concentration of counterions associated
with the negatively charged atomic groups of DNA macromolecule.
The first term in (\ref{model8}) describes the contribution from
DNA counterions to the conductivity of the system. The second term
in (\ref{model8}) describes the conductivity loss caused by the
association of counterions with the phosphate groups of DNA
macromolecule. Taking into account the formulae (\ref{model6}),
(\ref{model7}) and (\ref{model8}) the contribution of DNA to the
conductivity of polyelectrolyte solution (\(\Delta \sigma =
\sigma_\text{DNA+salt} - \sigma_\text{salt}\)) may be determined
as follows:
\begin{equation}
\Delta \sigma = c_\text{p} \lambda^{+} - A_\text{2}(c) \lambda^{+}
- A_\text{1}(c)( \lambda^{+} + \lambda^{-}). \label{model9}
\end{equation}

The concentration of condensed ions may be considered proportional
to the concentration of salt and concentration of DNA phosphate
groups, respectively: \(A_\text{1}(c) = \beta (c)c\) and
\(A_\text{2}(c) = \alpha (c)c_\text{p}\). The coefficients
\(\alpha (c)\) and \(\beta (c)\) depend on concentration of added
salt and describe the part of the ions condensed on macromolecule
surface. Let us consider the functions \(\alpha (c)\) and \(\beta
(c)\) in linear approximation:
\begin{equation}
\alpha (c) = \alpha_\text{0} + \alpha_\text{1} c; \quad \beta (c)
= \beta_\text{0} + \beta_\text{1} c, \label{model10}
\end{equation}
where \(\alpha_\text{0}\), \(\alpha_\text{1}\), \(\beta_\text{0}\)
and \(\beta_\text{1}\) are the parameters that may be determined
from the following conditions.

In case of salt free solution (\(c\) = 0) the conductivity is
determined by free counterions of DNA and $\alpha|_\text{c=0}$ =
0, thus $\alpha_{0}=0$. Increasing salt concentration the degree
of neutralization of DNA surface increases, and under some
concentration point (\(c = c_\text{cr}\)) all phosphate groups of
the double helix become neutralized. Since the counterions,
attached to DNA macromolecule, are not involved in the
conductivity process the condition $ \alpha |_{c \geq c_{cr}}$=1
should be valid, thus $\alpha_{1}=1/c_{cr}$. The ions of added
salt condense on counterions that are already tethered to the
phosphate groups of DNA backbone, therefore $\beta|_{c \leq
c_{cr}}$ = 0, and $\beta_{0}=-\beta_{1}c_{cr}$. Further increase
of salt concentration leads to the crystallization of salt ions,
and under some defined concentration (\(c = c_\text{max}\)) the
crystallization will be maximal that corresponds to the condition
$\beta|_{c = c_{max}}$ = 1, and $\beta_{1}=1/(c_{max}-c_{cr})$.
Taking into consideration these conditions the formulae
(\ref{model10}) may be written in the following form:
\begin{equation}
\alpha (c) = \frac{c}{c_\text{cr}}; \quad \beta (c) = \frac{c -
c_\text{cr}}{c_\text{max} - c_\text{cr}}. \label{model11}
\end{equation}

The temperature dependence of ion mobility may be taken into
consideration analogically to the equation (\ref{sigma4}):
\(\lambda = \lambda_\text{0} \exp(-\Delta E / k_\text{B} T)\),
where \(\lambda_\text{0}\) is the characteristic equivalent
mobility. The value of \(\lambda_\text{0}\) may be determined
using known values of ion mobility for some defined temperature
\(T_\text{0}\): \(\lambda_\text{0} = \lambda
(T_\text{0})\exp(\Delta E / k_\text{B} T_\text{0}) \). Taking this
into consideration, and substituting the formulae (\ref{model11})
to the equation (\ref{model9}), the formula for \(\Delta \sigma\)
may be written in the following form:

 \[\Delta\sigma=\left\{\begin{array}{l}
{\frac{(c_{cr}-c)c_{p}\lambda_{0}^{+}}{c_{cr}}
\exp\left[-\frac{\Delta
E(1-T/T_{0})}{k_{B}T}\right], \quad c\leq c_{cr};}\\
{-\frac{(c-c_{cr})c(\lambda_{0}^{+}+\lambda_{0}^{-})}{c_{max}-c_{cr}}
\exp\left[-\frac{\Delta E(1-T/T_{0})}{k_{B}T}\right], \quad c >
c_{cr}.}
\end{array}\right. \]
In the equation (12) \(\lambda_\text{0} ^{+}\) and
\(\lambda_\text{0} ^{-}\) are the mobility of positively and
negatively charged ions under characteristic temperature
\(T_\text{0}\).

It is seen that \(\Delta \sigma\) values are positive under the
salt concentration range \(c \leq c_\text{cr}\). In case of high
concentrations of added salt (\(c>c_\text{cr} \)) the values
\(\Delta \sigma \) are negative. The contribution of DNA to the
conductivity of polyelectrolyte is inessential (\(\Delta
\sigma\)~=~0) when all phosphate groups of DNA backbone are
neutralized (\(c = c_\text{cr} \)). Note the developed model does
not take into consideration the degradation process of DNA
macromolecules under the melting temperatures.

\section{Discussion}\label{disc}
\begin{figure}
\begin{center}
\resizebox{0.55\textwidth}{!}{%
  \includegraphics{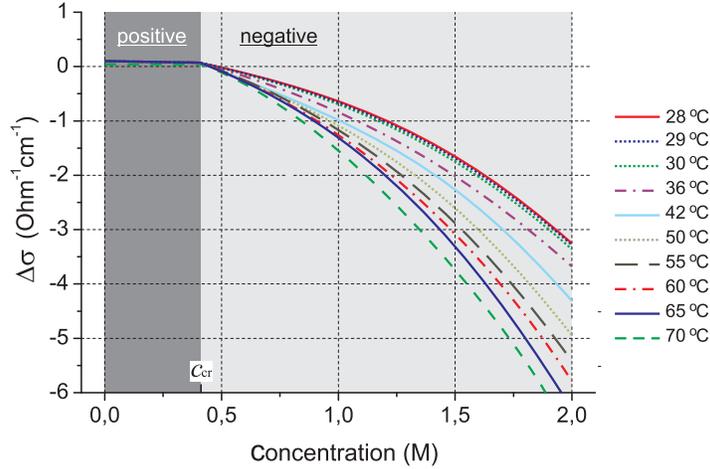}
}
\end{center}
 \caption{Concentration dependence for the difference between
conductivity of DNA and of electrolyte solution on salt
concentration, calculated by the formula (12).}
\label{fig5}       % Give a unique label
\end{figure}

To characterize the influence of DNA macromolecules on
conductivity of the system let us estimate \(\Delta \sigma\) by
formula (12). The parameters, necessary for the calculations, are
determined as follows. The concentration of phosphate groups in
solution is determined according to the concentration of DNA in
the experimental samples (2~mg/ml) \(c_\text{p} \)~=~6.35~M. The
maximal salt concentration is taken the same as solubility limit
of KCl  \(c_\text{max} \)~=~4.6~M \cite{40}. The value of critical
concentration of added salt \(c_\text{cr}\)~=~0.4~M is determined
from the condition \(\Delta \sigma\)~=~0. The characteristic
mobility \(\lambda_\text{0} ^{+}\) and \(\lambda_\text{0} ^{-}\)
for K$^{+}$ and Cl$^{-}$ ions are taken the same as in electrolyte
solution \(\lambda_\text{0} ^{+}\)~=~55.1
cm$^{2}\Omega^{-1}$mole$^{-1}$ and \(\lambda_\text{0}
^{-}\)~=~55.8 cm$^{2}\Omega^{-1}$mole$^{-1}$ under the temperature
25~$^{0}$C \cite{37}. The potential barrier \(\Delta E \approx
16\) kJ/mole is taken as average value of activation energies
(Table~\ref{barrier}). As the result the concentration dependences
of \(\Delta \sigma\) are shown in Figure~\ref{fig5}.

It is seen that the conductivity of DNA solution in concentration
range \(c < c_\text{cr}\) is practically the same as the
conductivity of respective electrolyte solution, and
\(\Delta\sigma \) is positive. At higher concentrations (\(c >
c_\text{cr}\)) the obtained difference between conductivity of DNA
solution and electrolyte solution is negative. Increasing the
temperature, the values of $\Delta \sigma$ decrease in this
concentration range. The calculated results (Figure~\ref{fig5})
qualitatively agree with the experimental data
(Figure~\ref{fig4}). However, in the concentration range \(c <
c_\text{cr}\) the experimentally observed values of
\(\Delta\sigma\) are larger, which may be caused by complexity of
the mechanism of counterion condensation on DNA.

\begin{figure}[t!]
\begin{center}
\resizebox{0.65\textwidth}{!}{%
  \includegraphics{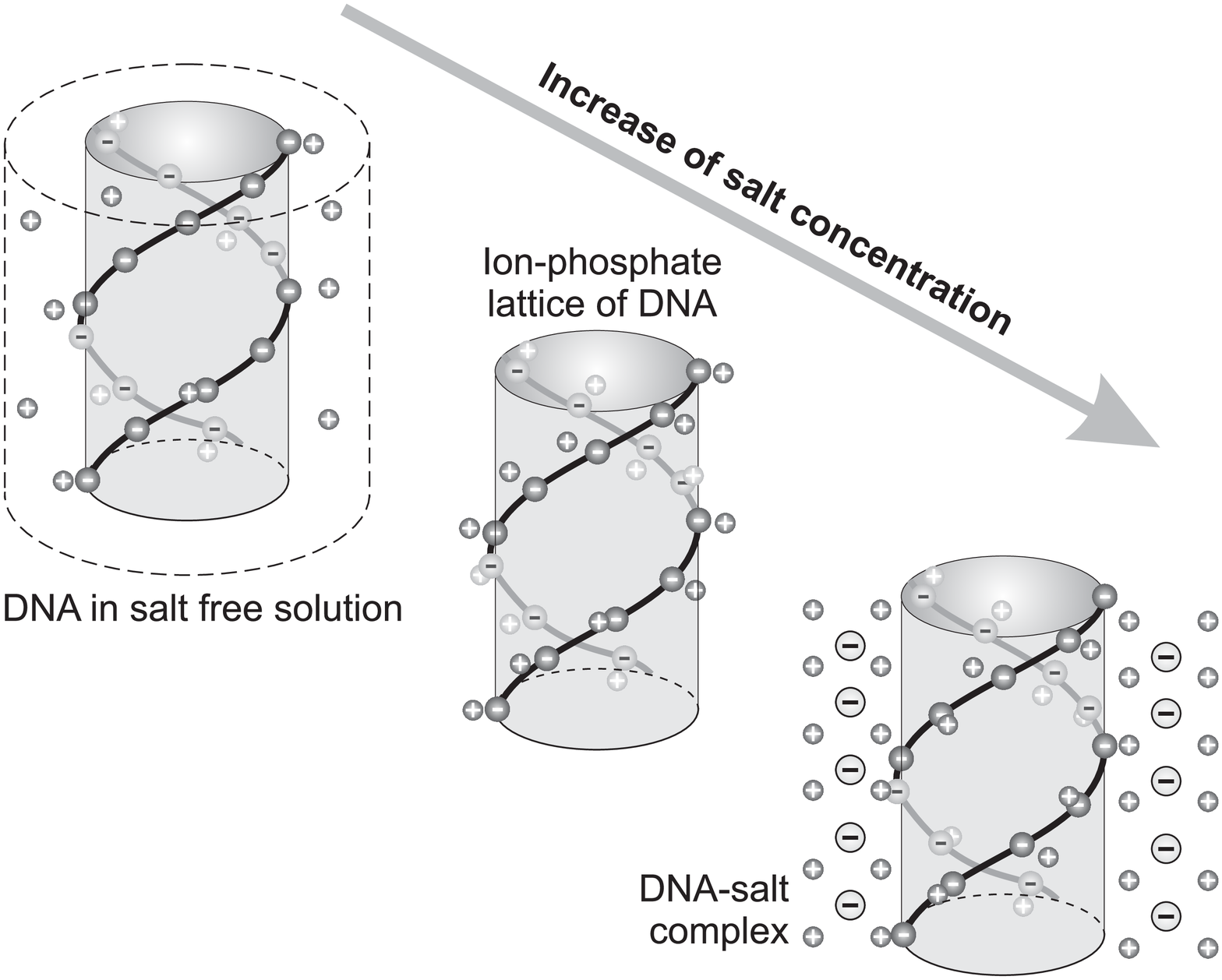}
}
\end{center}
 \caption{Scheme of the process of ionic structuring around DNA double helix at
different concentrations of added salt.}
\label{fig6}       % Give a unique label
\end{figure}

According to the results of estimations the following mechanism of
counterion ordering around DNA macromolecules may be introduced.
Under the low concentration of added salt the degree of phosphate
group neutralization is about the same as in the case of salt free
solution (Figure~\ref{fig6}a). The counterions come off the
ion-hydrate shell of macromolecule and determine the conductivity
of the system. Increasing salt concentration the number of
neutralized phosphate groups increases and under the critical
concentration the phosphate groups should be completely
neutralized (Figure~\ref{fig6}b). The counterions with the
phosphate groups form electrically neutral system resembling to
the lattice of ionic crystal (ion-phosphate lattice)
\cite{30,31,32,33}. The formation of DNA ion-phosphate lattice
induces the decrease of conductivity of the system. After the
formation of ion-phosphate lattice salt ions condense on
counterions tethered to the phosphate groups of macromolecule, and
DNA-salt complexes are formed (Figure~\ref{fig6}c). Such complexes
may be observed as the textures on a surface after evaporation of
solution \cite{16}. Formation of DNA-salt complexes reduces the
conductivity of the system due to the decrease of the number of
positively and negatively charged ions involved in the electric
current.

\section{Conclusions}
\label{conclusion} In the present work the ordering of ions in DNA
water solutions is studied by conductivity experiments. As the
result the temperature dependence (from 28 to 70~$^{0}$C) of
conductivity for DNA solution with KCl salt (the concentration
from 0 to 2~M) are obtained. In case of salt free solution there
exist three characteristic temperature ranges describing the
stages of the melting process of DNA double helix. In case of DNA
with added salt the characteristic stages of DNA melting are
hardly distinguishable that may be due to the stabilization of the
double helix by the ions of added salt. The comparison between
conductivity of DNA solution with the added salt and electrolyte
solution shows that under the concentrations lower than 0.4~M
(critical concentration) the conductivity of DNA solution is
higher than the conductivity of respective electrolyte. Starting
from the critical concentration the conductivity of electrolyte is
higher than the conductivity of DNA solution.

Basing on developed phenomenological model for the conductivity of
DNA solution, the mechanism of ionic ordering in DNA solution is
introduced. It is considered that under the low concentrations of
added salt DNA counterions do essential contribution to the
electrical conductivity of the system. Increasing salt
concentration to the critical one the counterions condense on DNA
macromolecule and the ion-phosphate lattice is formed. Further
increase of salt concentration leads to the condensation of anions
on cations attached to the phosphate groups of DNA backbone and
DNA-salt complexes are formed. Growth of the DNA-salt complexes
decreases the conductivity of the system. The introduced mechanism
qualitatively describes the experimentally observed changes of
conductivity of DNA solutions.

\section{Acknowledgements}
The present work was partially supported by the Program of
Fundamental Research of the Department of Physics and Astronomy of
the National Academy of Sciences of Ukraine (project No.
0112U000056) and by the State Fund for Fundamental Researches of
Ukraine (project No. 0112U007406). We thank Dr.~S.Ya.~Mandryk for
consultation on preparation of DNA samples.

\begin{flushright}
%{\footnotesize}
\end{flushright}

\end{document}